\documentclass[%
preprint,
showkeywords,
 amsmath,amssymb,
 aps,
]{revtex4-1}
\usepackage{amsmath,lipsum}
\usepackage{graphicx}
\usepackage{dcolumn}
\usepackage{bm}

\usepackage{mathrsfs}  
\usepackage[export]{adjustbox}

\usepackage{float}

\usepackage{subfloat}
\usepackage{subfig}

\usepackage{caption}

\newcommand{\bra}[1]{\langle #1 |}
\newcommand{\ket}[1]{| #1 \rangle}

\newcommand{\wigmtx}[4]{\mathscr{D}_{#2 \, #3}^{\,#1}\left(#4\right)}
\newcommand{\wjm}[6]{ \left( \begin{array}{c c c} #1 & #2 & #3 \\ #4 & #5 & #6 \end{array} \right)}

\begin{document}


\title{Pair-eigenstates and mutual alignment \\ of coupled molecular rotors in a magnetic field}

\author{Ketan Sharma}
 \email{ketan@fhi-berlin.mpg.de}
\author{Bretislav Friedrich}
 \email{bretislav.friedrich@fhi-berlin.mpg.de}
\affiliation{%
 Fritz-Haber-Institut der Max-Planck-Gesellschaft\\
 Faradayweg 4-6, D-14195 Berlin, Germany
}%

\date{\today}

\begin{abstract}
We examine the rotational states of a pair of polar $^2\Sigma$ molecules subject to a uniform magnetic field. The electric dipole-dipole interaction between the molecules creates entangled pair-eigenstates of two types. In one type, the Zeeman interaction between the inherently paramagnetic molecules and the magnetic field destroys the entanglement of the pair-eigenstates, whereas in the other type it does not. The pair-eigenstates exhibit numerous intersections, which become avoided for pair-eigenstates comprised of individual states that meet the selection rules  $\Delta J_{i}=0,\pm 1$, $\Delta N_{i}=0,\pm 2$, and $\Delta M_{i}=0,\pm 1$  imposed by the electric dipole-dipole operator. Here $J_{i}$, $N_{i}$ and $M_{i}$ are the total, rotational and projection angular momentum quantum numbers of molecules $i=1,2$  in the absence of the electric dipole-dipole interaction. We evaluate the mutual alignment of the pair-eigenstates and find it to be independent of the magnetic field, except for states that undergo avoided crossings, in which case the alignment of the interacting states is interchanged at the magnetic field corresponding to the crossing point. We present an analytic model which provides ready estimates of the pairwise alignment cosine that characterises the mutual alignment of the coupled rotors.

\end{abstract}

\pacs{Valid PACS appear here}
\keywords{Alignment, Zeeman effect, Entanglement, Bell states, Polar Paramagnetic Molecules}

\maketitle


\section{Introduction}
 
 External electric, magnetic and optical fields can be used to manipulate not only the rotational \cite{SlenFriHerPRL1994, FriSlenHer1994, JCP1999FriHer,  FriHerJPCA99, Ortigoso1999, Seideman:99a, Seideman1999, PCCP2000FriHer, JCP2000BoFri, Larsen:00a, CaiFri2001, CCCC2001, Averbukh:01a, Leibscher:03a, PRL2003Sakai, JMO2003-Fri, Frie2003-ModernTrends, Leibscher:04a, Buck-Farnik, HaerteltFriedrichJCP08, PotFarBuckFri2008, Leibscher:09a, PRL2009Stap, Owschimikow2009, Ohshima2010, Krems2010, Averbukh2010, Owschimikow2010, Owschimikow2011, StapPRL2012,Sharmafriedrich2015} and translational \cite{ZhaoFriChung2003,doyle2004editorial, KremsYe2009, Demilleetal2005, Southwell2002, Bethlem2000, e4,Shuman2010, Barry2012, Zeppenfeld2012, ChemRev2012-Meijer,Barry2014, grimm_et_al, Bloch_ManyBodyGases,LemKreDoyKais:MP2013} motion of individual molecules but  also to modify and engineer intermolecular potentials \cite{Zoller2007,LemeshkoPrA2011,FriedrichMP2012}. This is of relevance to few- and many-body physics where the ability to manipulate intermolecular potentials can be harnessed to, for instance,  engineer new phases \cite{Buechler-Demler-Lukin-Micheli-Prokofev-Pupillo-Zoller2007,Micheli-Pupillo-Buechler-Zoller2007}, implement Hubbard-type Hamiltonians with controllable parameters \cite{Buechler-Micheli-Zoller2007}, simulate spin models \cite{Micheli-Brennen-Zoller2006}, or realise  the dissipative bond \cite{Lemeshko-Weimer2013, Lemeshko2013}. In our recent work \cite{LemeshkoPrA2011,FriedrichMP2012}, we presented a method for manipulating the interaction potential between a pair of polar $^1\Sigma$ molecules with far-off-resonant light. That method is based on the triple-combination of the electric dipole-dipole, anisotropic polarisability, and the retarded induced dipole-dipole interactions and offers a wide tunability range of the intermolecular potentials that it generates.  

Herein, we examine how the electric dipole-dipole interaction potential between two polar  $^2\Sigma$ molecules -- which are  inherently paramagnetic -- creates entangled pair-eigenstates and how these are affected by the Zeeman  interaction between the molecules and the magnetic field. The electric dipole-dipole intermolecular potential couples Zeeman levels that fulfil selection rules imposed by the electric dipole-dipole operator. This coupling alters the Zeeman levels of the pair-eigenstates in general and modifies the mutual alignment of the two molecular rotors in particular. We are reminded of the coupling of the Zeeman levels of a single polar paramagnetic molecule by a superimposed electric field \cite{PCCP2000FriHer,JCP2000BoFri}, whose interaction with the body-fixed electric dipole of the polar molecule plays the role of the electric dipole-dipole interaction (although under different selection rules). However, the pair-eigenstates  exhibit a  behaviour quite different from that of single-molecule eigenstates. For instance, we find that the field-free pair-eigenstates are the maximally entangled Bell states \cite{Mermin1993}. The application of a magnetic field is akin to effecting a Bell measurement that results in destroying the pair's entanglement. These features predestine such pair-eigenstates to be employed as qubits in a quantum computation scheme based on an array of trapped $^2\Sigma$ molecules \cite{KarSharFri2016}. Previous proposals relied on the Stark states of trapped polar linear \cite{DeMille2002,FriHerKais2011a,FriHerKais2013} and symmetric top \cite{FriHerKais2011b} molecules as qubits. 
  
This paper is organised as follows. In Section \ref{sec:theory}, we present the basic theory of the interaction of a pair of polar $^2\Sigma$ molecules with a magnetic field, starting with a single such molecule in Subsection \ref{sec:theory1} and laying out the full-fledged theory for the two-molecule system in Subsection \ref{sec:theory2}. In section \ref{sec:rnd}, we present and discuss our results on the two-molecule system in the absence (Subsection \ref{sec:without}) and presence of a weak (Subsection \ref{sec:weak}) and strong (Subsection \ref{sec:strong}) electric dipole-dipole coupling as a function of the magnetic field strength. In Subsection \ref{sec:alignment} we present and discuss our results on the mutual alignment of the two molecules and in Subsection \ref{sec:analm} we introduce a model for evaluating the mutual alignment of the two coupled molecular rotors. Section \ref{sec:conc} provides a summary of our results. Appendices \ref{app:dip_dip} and \ref{app:pwelements} show derivations of the matrix elements of the electric dipole-dipole operator and the pairwise alignment cosine in the cross-product basis set of the two molecules.

\section{Theory}
\label{sec:theory}
\subsection{The Hamiltonian of a polar $^2\Sigma$ molecule in a magnetic field}
\label{sec:theory1}

\begin{figure}[htp!]
 \centering
 \includegraphics[width=\textwidth, height=\textheight, keepaspectratio,frame]{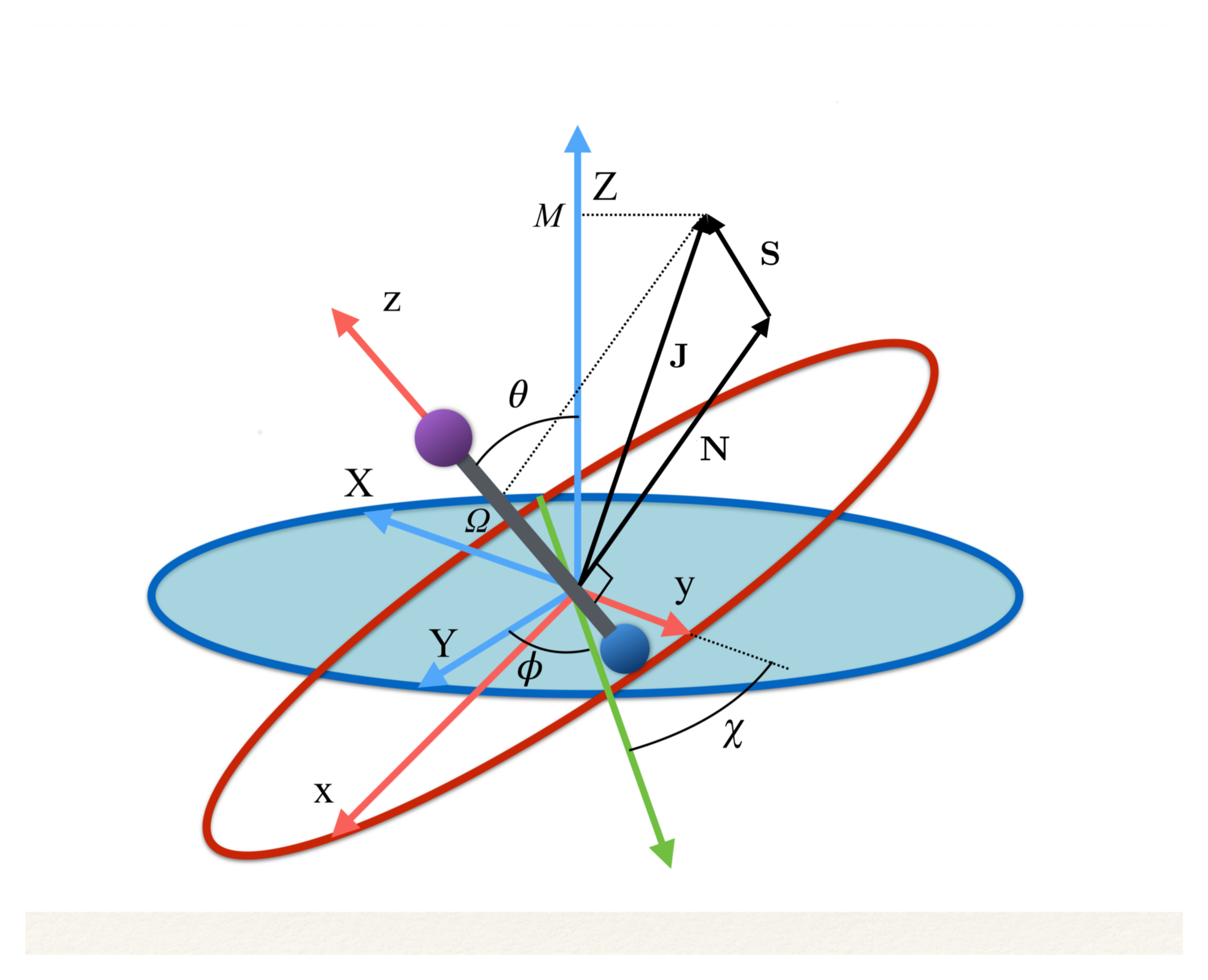}
\caption{Euler angles $(\phi, \theta, \chi)$ describing the rotation of the molecular coordinates $(x, y, z)$ fixed to a diatomic molecule (depicted as a bar-bell) with respect to the space-fixed coordinates $(X, Y, Z)$. The green axis is the line of nodes, perpendicular to both $z$ and $Z$. Also shown are the rotational, $\bf N$, electron spin, $\bf S$, and total, $\bf J$, angular momenta as well as the projections $M$ and $\Omega$ of $\bf J$ on the $z$  and $Z$ axes, respectively.}
\label{fig:DM1}
\end{figure}

We first consider an individual polar $^2\Sigma$ molecule in a uniform magnetic (Zeeman) field. Its Hamiltonian (apart from nuclear spin) $H$  is given by the sum of the rotational and Zeeman terms \citep{PCCP2000FriHer,Levefre-Field:2004,bunker-jensen2005,zare2013angular,Sharmafriedrich2015}.  
\begin{equation}
\label{eqn:hami}
H=B\textbf{N}^{2} + \gamma\textbf{N} \cdot \textbf{S} +B\eta_mS_Z
\end{equation}
where $B$ is the rotational constant, $\textbf{N}$ the rotational angular momentum operator, $\textbf{S}$ the electronic spin angular momentum operator, $\gamma$  the spin-rotation coupling constant and $S_Z$ the space-fixed $Z$ component of the molecule's electronic spin. The dimensionless magnetic interaction parameter is given by
\begin{equation}
\eta_m\equiv\frac{\mu_m \mathcal{H}}{B}
\label{etam}
\end{equation}
where $\mu_m=g_S\mu_B$ is the electronic magnetic dipole moment of the $^2\Sigma$ molecule, $g_S \cong 2.0023$ the electronic gyromagnetic ratio,  $\mu_B$ the Bohr magneton and $\mathcal{H}$ the magnetic field strength. 

Figure \ref{fig:DM1} shows the body- and space-fixed frames of reference $(x, y, z)$  and $(X, Y ,Z)$, respectively, along with the Euler angles $(\phi, \theta, \chi)$ that describe their mutual  rotation. The angular momenta ${\bf N}$  (rotational), ${\bf J}$ (total) and ${\bf S}$ (electron spin) are also shown, along the projections $M$ and $\Omega$ of the total angular momentum ${\bf J}$ on the space fixed $Z$-axis and molecule fixed $z$-axis, respectively. Note that ${\bf N}={\bf J}-{\bf S}$. 

While for a $^2\Sigma$ state the electronic spin angular momentum $S=\frac{1}{2}$, the orbital electronic angular momentum is identically zero and so is the spin-orbit coupling. A field-free $^2\Sigma$ state thus exhibits a Hund's case (b) coupling between the rotational and electronic angular momenta \cite{Levefre-Field:2004}, with  the projections of the total and spin electronic angular momenta on the molecular axis (an axis of cylindrical symmetry) $\Omega=\Sigma=\frac{1}{2}$, cf. Fig. \ref{fig:DM1}.

The Hund's case (b) basis functions are an equally weighted linear combination of Hund's case (a) basis functions, each a  product of a symmetric top wave function, 
\begin{equation}
\left| J,\Omega,M \right\rangle=(-1)^{M-\Omega}\sqrt{\frac{(2J+1)}{8\pi}}\mathfrak{D}^{J}_{-M,-\Omega}(\theta, \phi, \chi)
\end{equation}
and a spin function, 
\begin{equation}
\left| S, \Sigma \right\rangle= \frac{\alpha^{S+\Sigma}\beta^{S-\Sigma}}{\sqrt{(S+\Sigma)!(S-\Sigma)!}}
\end{equation}
with $J=N\pm S$ the total (rotation and electron spin) angular momentum quantum number, $M$ and $\Omega$ the projections of the total angular momentum on, respectively, the space-fixed $Z$ axis and the body-fixed $z$ axis,  $\mathfrak{D}^{J}_{M,\Omega}(\theta, \phi, \chi)$ the Wigner matrix, with $\theta, \phi, \chi$ the Euler angles, and $\alpha, \beta$ the spin functions. Thus for a  field-free $^2\Sigma$ state ($S=\frac{1}{2}$), there are two types of Hund's case (b) basis functions
\begin{equation}
\psi_\pm (N\pm\frac{1}{2},M)= \frac{1}{\sqrt{2}}\left[|S,\frac{1}{2}\rangle|J,\Omega,M\rangle\pm|S,-\frac{1}{2}\rangle|J,-\Omega,M\rangle\right]\equiv|N,J,M\rangle
\end{equation}
pertaining to $J=N\pm\frac{1}{2}$, with parity $(-1)^N$. The corresponding eigenenergies are
\begin{equation}
  E_+(N+\frac{1}{2}, M)  =  BN(N+1)+\frac{\gamma}{2}N 
   \label{eqn:fieldfree-a}
   \end{equation}
\begin{equation}
   E_-(N-\frac{1}{2}, M)  =  BN(N+1)-\frac{\gamma}{2}(N+1)
  \label{eqn:fieldfree-b}
\end{equation}
We note that both $J$ and $N$ but not $\Omega$ are good quantum numbers for a field-free $^2\Sigma$ molecule.

The $S_Z$ operator couples Hund's case (b)  basis functions with same $M$ but with $N's$ that are either the same or differ by $\pm2$ and hence have the same parity. The selection rule on $N$ moreover ensures that the Hamiltonian matrix in the Hund's case (b) basis for the Zeeman interaction of a $^2\Sigma$ molecule factors into blocks that are no greater than $2\times 2$, rendering the corresponding Zeeman energy at most quadratic in $\mathcal{H}$. 

The Zeeman states $\left | \tilde{N}, \tilde{J}, M; \eta_m \right\rangle $  of a $^2\Sigma$ molecule adiabatically correlate with the field-free rotor states $\left| N, J, M \right\rangle$ such that  $\left|\tilde{N}, \tilde{J}, M; \eta_m \rightarrow 0 \right\rangle \rightarrow \left| N, J, M \right\rangle$, where $\tilde{N}$ and $\tilde{J}$ are adiabatic labels rather than quantum numbers. The projection quantum number  $M$ and the parity $(-1)^{\tilde{N}}$ remain good quantum number even in the presence of the Zeeman field. The effects of the magnetic field on $^2\Sigma$ molecules have been discussed in greater detail, e.g., in Refs. \cite{PCCP2000FriHer,Sharmafriedrich2015}.

\subsection{The Hamiltonian of a pair of polar $^2\Sigma$ molecule in a magnetic field}
\label{sec:theory2}
\begin{figure}[htp!]
 \centering
  \includegraphics[width=\textwidth, height=\textheight, keepaspectratio, frame]{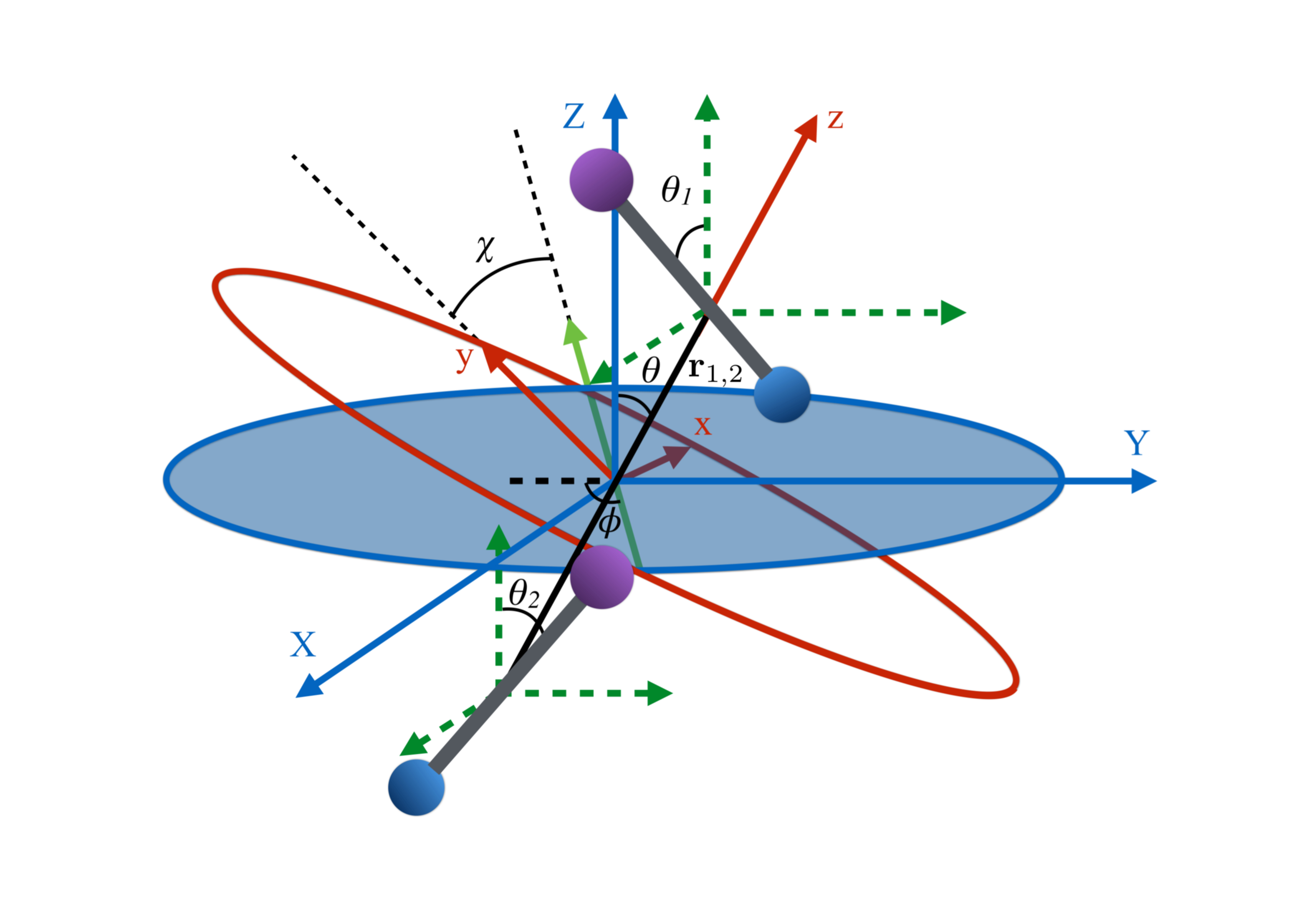}
  \caption{Definition of Euler angles $(\phi, \theta, \chi)$ describing the rotation of the intermolecular co-ordinate $(x, y, z)$ with respect to the space-fixed coordinates $(X, Y, Z)$ for two diatomic molecules depicted as a bar-bells. The intermolecular frame of reference has its $z$-axis aligned along the internuclear axis, ${{\bf r}_{1,2}}$. The green dashed coordinates are the space fixed coordinates $(X,Y,Z)$ translated to each molecule. The Euler angles for each molecule introduced  in Fig. \ref{fig:DM1} are from here on represented using subscripts 1 and 2 for each molecule.}
  \label{fig:twomolsys}
\end{figure}

We now consider a pair of polar $^2\Sigma$ molecules in the presence of a uniform magnetic field. The Hamiltonian of such a composite, two-molecule system is the sum of the single-molecule  Hamiltonians, $H_i$, and the electric and magnetic dipole-dipole coupling terms. Upon neglecting the much weaker magnetic dipole-dipole interaction, the Hamiltonian takes the form
\begin{equation}
\label{eqn:Main}
H = \sum_{i=1}^2H_{i} + V_{d-d},
\end{equation}
where  $i=1,2$ and $V_{d-d}$ is the electric dipole-dipole interaction.

The two molecule system is shown in Figure \ref{fig:twomolsys}, along with the space- and body-fixed reference frames ($X, Y, Z$) and ($x,y,z$). While the $Z$ axis is defined by the direction of the magnetic field vector, the $z$-axis coincides with the intermolecular axis. The Euler angles ($\theta, \phi, \chi$) parametrize the rotation matrix which transforms between the laboratory (space-fixed) and intermolecular (body-fixed) frames \cite{Varshalovich2008}.
The rotations between the body-fixed frames of molecules 1 and 2 and the laboratory frame are described by Euler angles ($\theta_1, \phi_1, \chi_1$) and ($\theta_2, \phi_2, \chi_2$).

The  electric dipole-dipole interaction potential is given by \cite{Trinity1}
\begin{equation}
V_{d-d}=\frac{\boldsymbol{\mu}_1 \cdot \boldsymbol{\mu}_2-3(\boldsymbol{\mu}_1 \cdot {\bf n})(\boldsymbol{\mu}_2 \cdot {\bf n})}{4\pi\epsilon_0 r_{1,2}^3}
\label{eqn:Vdd}
\end{equation}
with $\boldsymbol{\mu}_1$ and $\boldsymbol{\mu}_2$ the electric dipole moments of the two molecules  and $\bold{r}_{1,2}$ the relative position vector of the centres of mass of the two molecules whose direction is given by the unit vector $\bold{n}\equiv \frac{\bold{r}_{1,2}}{r_{1,2}}$, and  $\epsilon_0$ is the permittivity of the vacuum. As usual, $r_{1,2} \equiv |{\bf r_{1,2}}|$ and $\mu_{1,2} \equiv |\boldsymbol{\mu}_{1,2}|$. Moreover, in our case, $\mu_{1}=\mu_{2}\equiv \mu$. 

Eq. (\ref{eqn:Vdd}) can be recast  in terms of the Wigner matrices $\wigmtx{l}{m}{0}{\phi, \theta, \chi}$:
\begin{equation}
\label{eqn:Xi}
 V_{d-d}=-\sqrt{6}~\overline{\Xi} \sum_{\nu\,\lambda}C(1,1,2;\nu,\lambda,\nu+\lambda)\wigmtx{1}{-\nu}{0}{\phi_1,\theta_1,\chi_1}\wigmtx{1}{-\lambda}{0}{\phi_2,\theta_2,\chi_2}\wigmtx{2}{\nu+\lambda}{0}{\phi,\theta,\chi}
\end{equation}
where $C(J_1,J_2,J_3;M_1,M_2,M_3)$ are the Clebsch-Gordan coeffcients, $J_1$ and $J_2$ the angular momentum quantum numbers of molecules 1 and 2, $M_1$ and $M_2$ the projection of the angular momenta of molecules 1 and 2 on the space fixed axis $Z$, $(\theta_1,\phi_1)$ and $(\theta_2,\phi_2)$ the rotational coordinates of molecules 1 and 2,  $(\theta,\phi)$ the spherical coordinates of their relative position vector ${\bf r_{1,2}}$, and 
\begin{equation}
\label{eqn:Xi2}
\overline{\Xi}\equiv\frac{\mu_1\mu_2}{4\pi\epsilon_0 r_{1,2}^3} 
\end{equation}
is a parameter that characterises the strength of the electric dipole-dipole interaction. The dimensionless parameter $\Xi\equiv\frac{\overline{\Xi}}{B}$ measures the strength of the electric dipole-dipole interaction in terms of the rotational constant.

\begin{figure}
\centering
\includegraphics[width=1\textwidth, height=0.5\textheight, keepaspectratio]{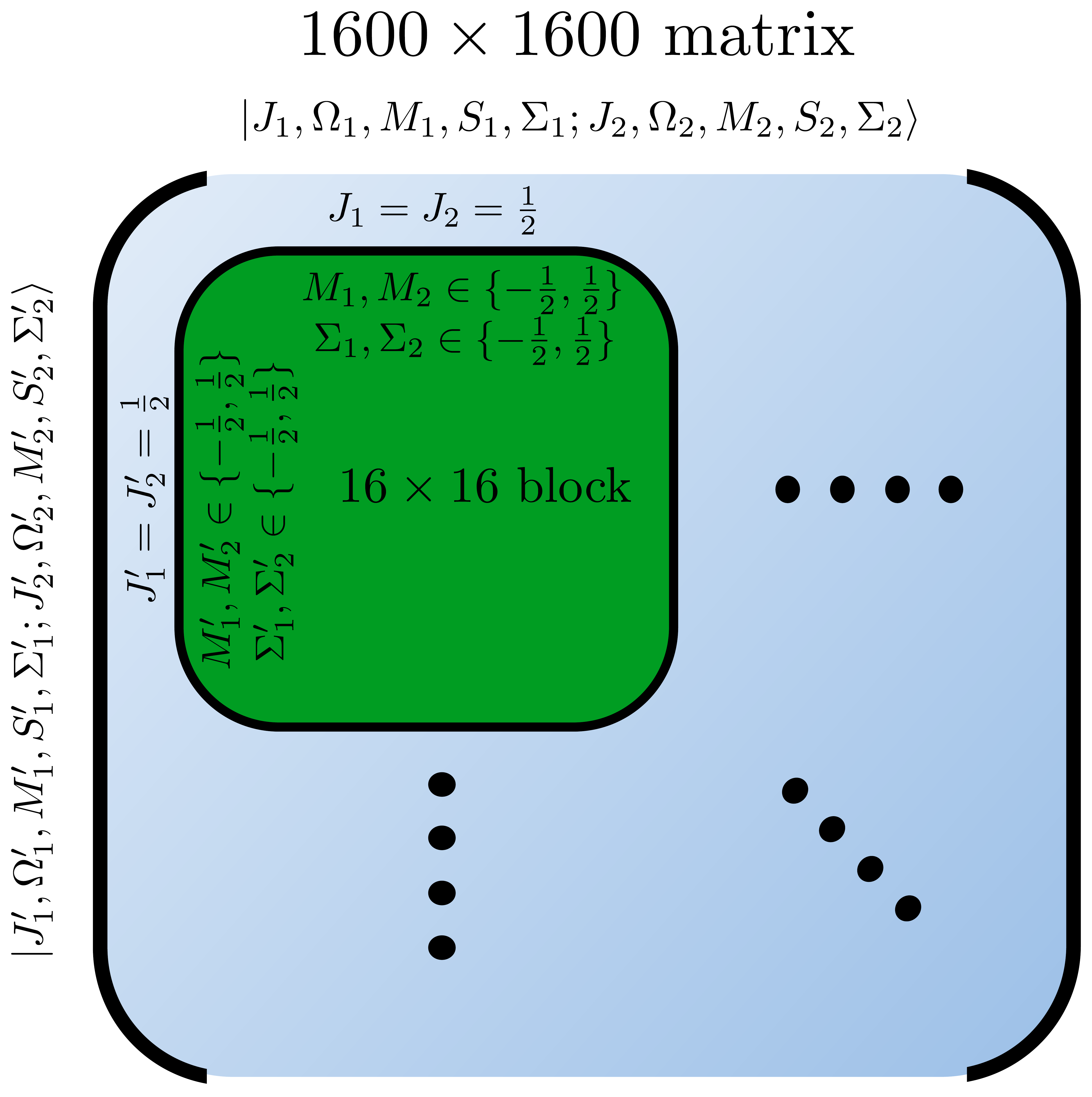}
\caption{Matrix representation of Hamiltonian of Eq. (\ref{eqn:Main}) in the cross product basis set $|J_1,\Omega_1,M_1,S_1,\Sigma_1;J_2,\Omega_2,M_2,S_2,\Sigma_2\rangle$ of two Hund's case (b) molecules, truncated such that $J_i$ with $i=1,2$ ranges from $\frac{1}{2}$ to $\frac{7}{2}$ for molecules $1$ and $2$. Hence $M_i$ ranges from $-J_i$ to $J_i$ while  $\Sigma_i=\pm\frac{1}{2}$. Same applies for primed quantities. Note that for instance $J_1=J_2=\frac{1}{2}=J_1^{'}=J_2^{'}$ give rise to a $16 \times 16$ sub-matrix. See text.}
\label{fig:TruncatedHamiltonianSchematic}
\end{figure}

The matrix elements of the Hamiltonian were calculated analytically in the cross product Hund's case (a) basis set, 
\begin{equation}\label{eqn:crossproduct}
  |J_1,\Omega_1,M_1,S_1,\Sigma_1;J_2,\Omega_2,M_2,S_2,\Sigma_2\rangle= \left|J_1 \Omega_1 M_1 \right\rangle\left| S_1 \Sigma_1 \right\rangle \otimes \left|J_2 \Omega_2 M_2 \right\rangle\left| S_2 \Sigma_2 \right\rangle
 \end{equation}
of the two molecules and the eigenproperties of the composite two-molecule system obtained by a numerical diagonalization of  a truncated Hamiltonian matrix, whose structure is shown in Figure \ref{fig:TruncatedHamiltonianSchematic}.

Note that the projection quantum numbers $\Omega_i$ and $\Sigma_i$ (with $i=1,2$) of the electronic angular momenta on the body-fixed axis of each $^2\Sigma$ molecule coincide, i.e., $\Omega_i=\Sigma_i$. The number of pairs of states determines the size of the basis set and is given by $[2\Sigma_{J_{\min}}^{J_{\max}} (2J+1)]^2$. For $J_{\min}=\frac{1}{2}$ and $J_{\max}=\frac{7}{2}$, this means that the truncated Hamiltonian matrix is of a $1600$ rank.

The matrix elements in the cross product of Hund's case (a) basis of the two molecules have been obtained in closed form, see Appendix \ref{app:dip_dip}:
\begin{equation}
 \begin{split}
  \bra{J_1^\prime \Omega_1^\prime M_1^\prime S_1^\prime \Sigma_1^\prime & J_2^\prime \Omega_2^\prime M_2^\prime S_2^\prime \Sigma_2^\prime}  V_{dd}\ket{J_1 \Omega_1 M_1 S_1 \Sigma_1 J_2 \Omega_2 M_2 S_2 \Sigma_2}  \\
  = & -\sqrt{30}\Xi B\left[2J_1'+1\right]^\frac{1}{2}\left[2J_1+1\right]^\frac{1}{2}\left[2J_2'+1\right]^\frac{1}{2}\left[2J_2+1\right]^\frac{1}{2}\\
  & \times \wjm{J_1'}{1}{J_1}{\Omega_1'}{0}{\Omega_1}\wjm{J_2'}{1}{J_2}{\Omega_2'}{0}{\Omega_2}   \delta_{S_1^\prime S_1}\delta_{S_2^\prime S_2} \delta_{\Sigma_1^\prime \Sigma_1} \delta_{\Sigma_2^\prime \Sigma_2}\\
    & \times \, \sum_{\nu\,\lambda} \wjm{1}{1}{2}{\nu}{\lambda}{-\nu-\lambda}\wigmtx{2}{\nu+\lambda}{0}{\phi,\theta,\chi}\wjm{J_1'}{1}{J_1}{M_1'}{-\nu}{M_1}\wjm{J_2'}{1}{J_2}{M_2'}{-\lambda}{M_2}    
 \end{split}
 \label{eqn:dip-dip}
\end{equation}
Eq. (\ref{eqn:dip-dip}) implies that the electric dipole-dipole interaction couples states with $\Delta M_1=0,\pm 1$, $\Delta J_1=0,\pm 1$, $\Delta M_2=0,\pm 1$ and $\Delta J_2=0, \pm 1$ of molecules 1 and 2. Thus, even in the absence of external fields, $M$ is not a good quantum number in the presence of the electric dipole-dipole interaction. In section \ref{sec:weak} we will introduce a labelling of states that circumvents this difficulty.
	
\section{Results and Discussion}
\label{sec:rnd} 
The diagonalization of the $1600\times1600$ Hamiltonian matrix was carried out using the Armadillo C++ linear algebra library \cite{Sanderson2010}. The states were  adiabatically tracked as a function of the magnetic field interaction parameter $\eta_m$ by monitoring the inner product between the eigenvector of a given state at the initial value of $\eta_m$ and all  possible eigenvectors for the new value of $\eta_m$. The calculations were carried out for the example of a NaO molecule, whose parameters are summarised in Table \ref{table:NaO}.  

\begin{table*}[!ht]
\centering
\caption{\small Rotational constant, $B$, spin-rotation constant, $\gamma$, electric dipole moment, $\mu$,  and values of the dimensionless interaction parameter $\eta _{m}$ at a magnetic field of 1 Tesla for  NaO(A$^2\Sigma$); also shown is the value of the dimensionless electric dipole-dipole interaction parameter $\Xi$, see Eq. (\ref{eqn:Xi2}). Compilation based on Refs. \cite{Worsnop-Herschbach} and our own calculations. $^a$Calculated using Gaussian 09.  $^b$Becke3LYP type calculation using TZP-DKH basis \cite{feller1996role,schuchardt2007basis}.}
\vspace{.3cm}
\label{table:NaO}
\begin{tabular}{| c | c | c | c | c | c | c | c |}
\hline 
\hline
$B$ [cm$^{-1}$] & $\gamma$ [cm$^{-1}$] & $\mu$ [D] &   $\eta_{m}$  @ 1 T & $\Xi$ @ 500 nm \\[5pt]

\hline

0.462 & 0.193 & 7.88$^{a,b}$  &  2.02  & $5.42 \times 10^{-6}$ \\[5pt]

 \hline
 \hline
  
\end{tabular}
\end{table*}

 \subsection{Pair-eigenstates in the absence of the  electric dipole-dipole coupling, $\Xi=0$}
 \label{sec:without}
 \begin{figure}[t!]
 \centering
  \includegraphics[width=\textwidth, height=0.7\textheight, keepaspectratio]{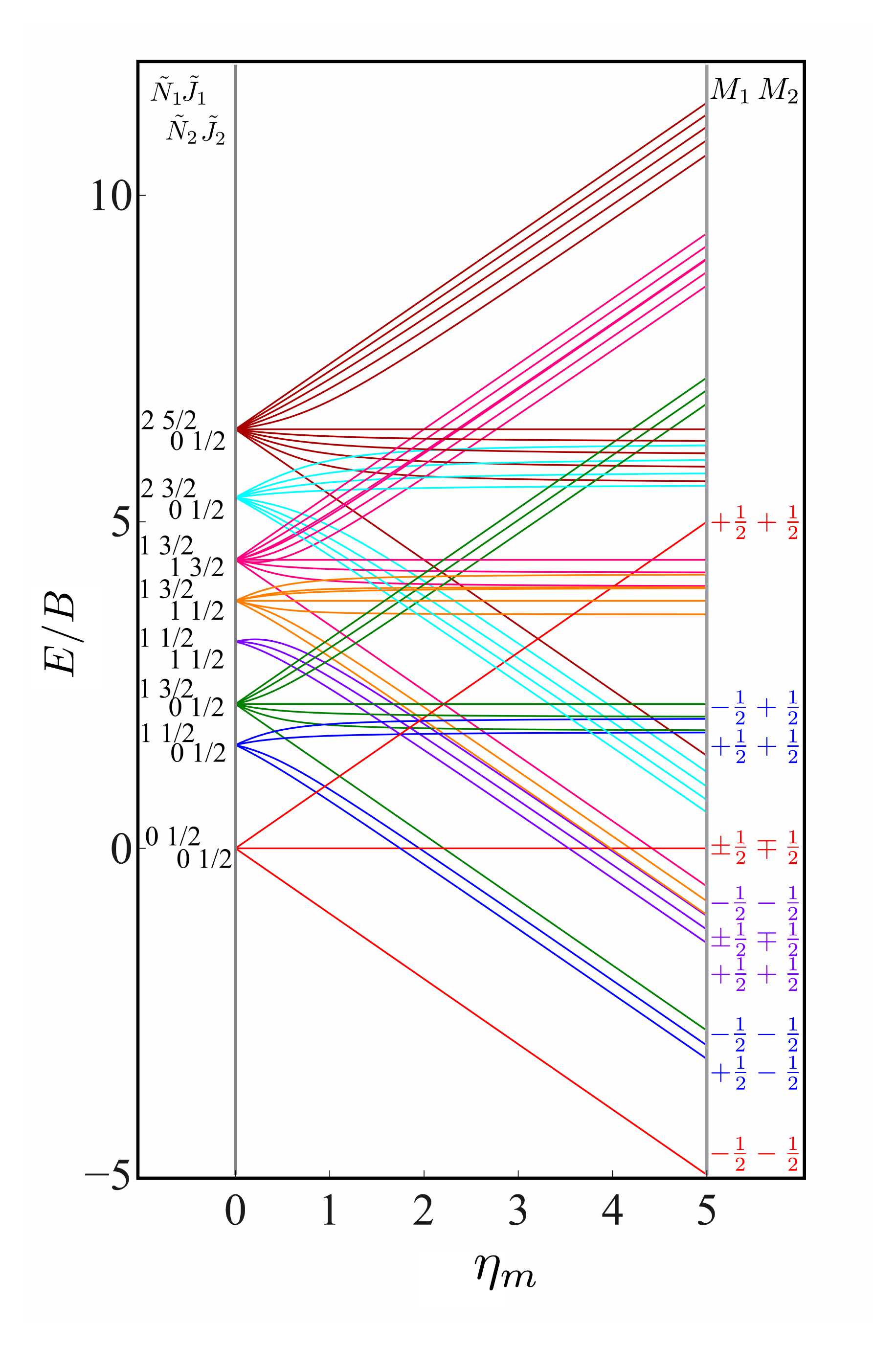}
  \caption{Dependence of the eigenenergies $E$ of the system of two polar paramagnetic $^2\Sigma$ molecules on the magnetic field strength parameter $\eta_m$ in the absence of the electric dipole-dipole interaction ($\Xi=0$). The eigenenergies are measured in terms of the rotational constant $B$.}
  \label{fig:novdd}
\end{figure}
 In the absence of the electric dipole-dipole interaction, i.e. for $\Xi=0$, the pair-eigenstates of the two-molecule system can be decomposed into products of eigenstates of the individual molecules,
\begin{equation}
|\tilde{J_1}, \tilde{N_1}, \tilde{M_1}; \tilde{J_2}, \tilde{N_2}, \tilde{M_2};\eta_m\rangle=|\tilde{J_1}, \tilde{N_1}, \tilde{M_1};\eta_m\rangle|\tilde{J_2}, \tilde{N_2}, \tilde{M_2};\eta_m\rangle
 \label{eqn:statedcmp}
\end{equation}
This implies that the two Hamiltonians $H_1$ and $H_2$, cf. Eqs. (\ref{eqn:hami}) and (\ref{eqn:Main}), can be diagonalised separately in order to obtain the eigenfunctions $|\tilde{J_1}, \tilde{N_1}, \tilde{M_1};\eta_m\rangle$ and $|\tilde{J_2}, \tilde{N_2}, \tilde{M_2};\eta_m\rangle$ and the corresponding eigenenergies $E_1$ and $E_2$. The eigenenergy of the two-molecule system is then calculated to be
 \begin{equation}
  E=E_1+E_2
 \end{equation}

Figure \ref{fig:novdd} shows the eigenenergies (in units of the rotational constant $B$) of the two-molecule system for $\Xi=0$. Each set of eigenstates with the same $\tilde{J}_1, \tilde{N}_1, \tilde{J}_2$ and $\tilde{N}_2$ is plotted in the same colour. The projection quantum numbers  $M_1$ and $M_2$ of the individual molecules are good quantum numbers in the absence of the electric dipole-dipole interaction.

 \subsection{Pair-eigenstates in the presence of a small dipole-dipole coupling, $\Xi\ll1$}
 \label{sec:weak}
  \begin{figure}[t!]
 \centering
  \includegraphics[width=\textwidth, height=0.7\textheight, keepaspectratio]{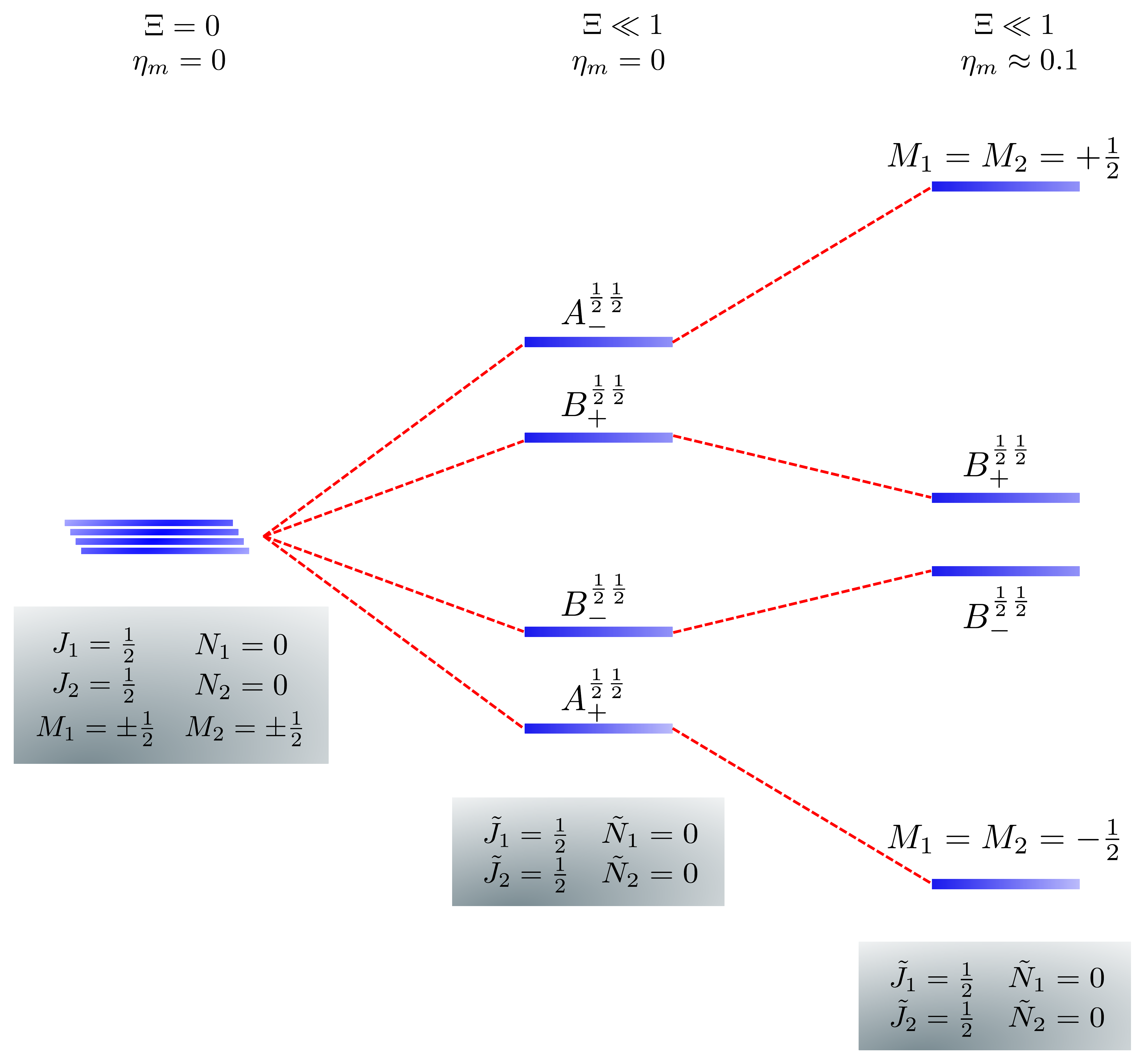}
  \caption{Correlation diagram involving the $\tilde{J}_1=\frac{1}{2}, \tilde{N}_1=0, \tilde{J}_2=\frac{1}{2}, \tilde{N}_2=0$ pair-eigenstates. The eigenstates, labelled in accordance with Table \ref{tbl:statelable}, are degenerate in the absence of the electric dipole-dipole interaction but their degeneracy is lifted when $\Xi\neq 0$.  The $A^{\frac{1}{2}\frac{1}{2}}_+$ and $A^{\frac{1}{2}\frac{1}{2}}_-$ states adiabatically transform into $M_1=M_2=-\frac{1}{2}$ and $M_1=M_2=\frac{1}{2}$ states, respectively, when the magnetic field is applied. Note that the $B$ states maintain their entanglement throughout.}
  \label{fig:cordiag1}
\end{figure}
  \begin{figure}[t!]
 \centering
  \includegraphics[width=\textwidth, height=0.7\textheight, keepaspectratio]{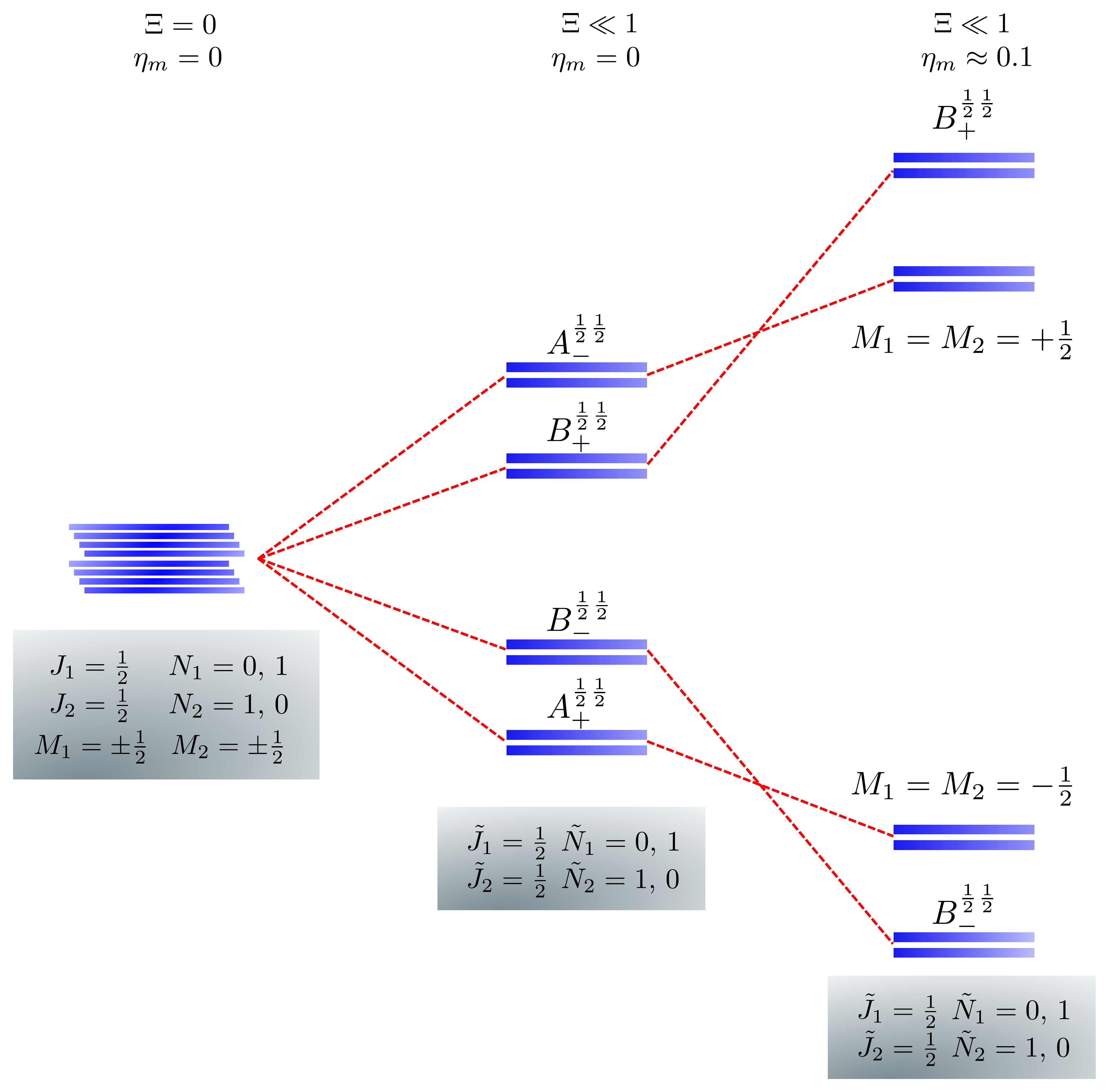}
  \caption{Correlation diagram involving the $\tilde{J}_1=\frac{1}{2}, \tilde{N}_1=0,1, \tilde{J}_2=\frac{1}{2}, \tilde{N}_2=1,0$ pair-eigenstates. The eigenstates, labelled in accordance with Table \ref{tbl:statelable},  are eight fold degenerate in the absence of electric dipole-dipole interaction but only doubly degenerate when $\Xi\neq 0$. This double degeneracy arises because the indistinguishability of two molecules. $A^{\frac{1}{2}\frac{1}{2}}_+$ and $A^{\frac{1}{2}\frac{1}{2}}_-$ states adiabatically transform into $M_1=M_2=-\frac{1}{2}$ and $M_1=M_2=\frac{1}{2}$ states, respectively, when the magnetic field is applied. Note that the $B$ states maintain their entanglement throughout.}
    \label{fig:cordiag2}
\end{figure}
  \begin{figure}[htp!]
 \centering
  \includegraphics[width=\textwidth, height=0.7\textheight, keepaspectratio]{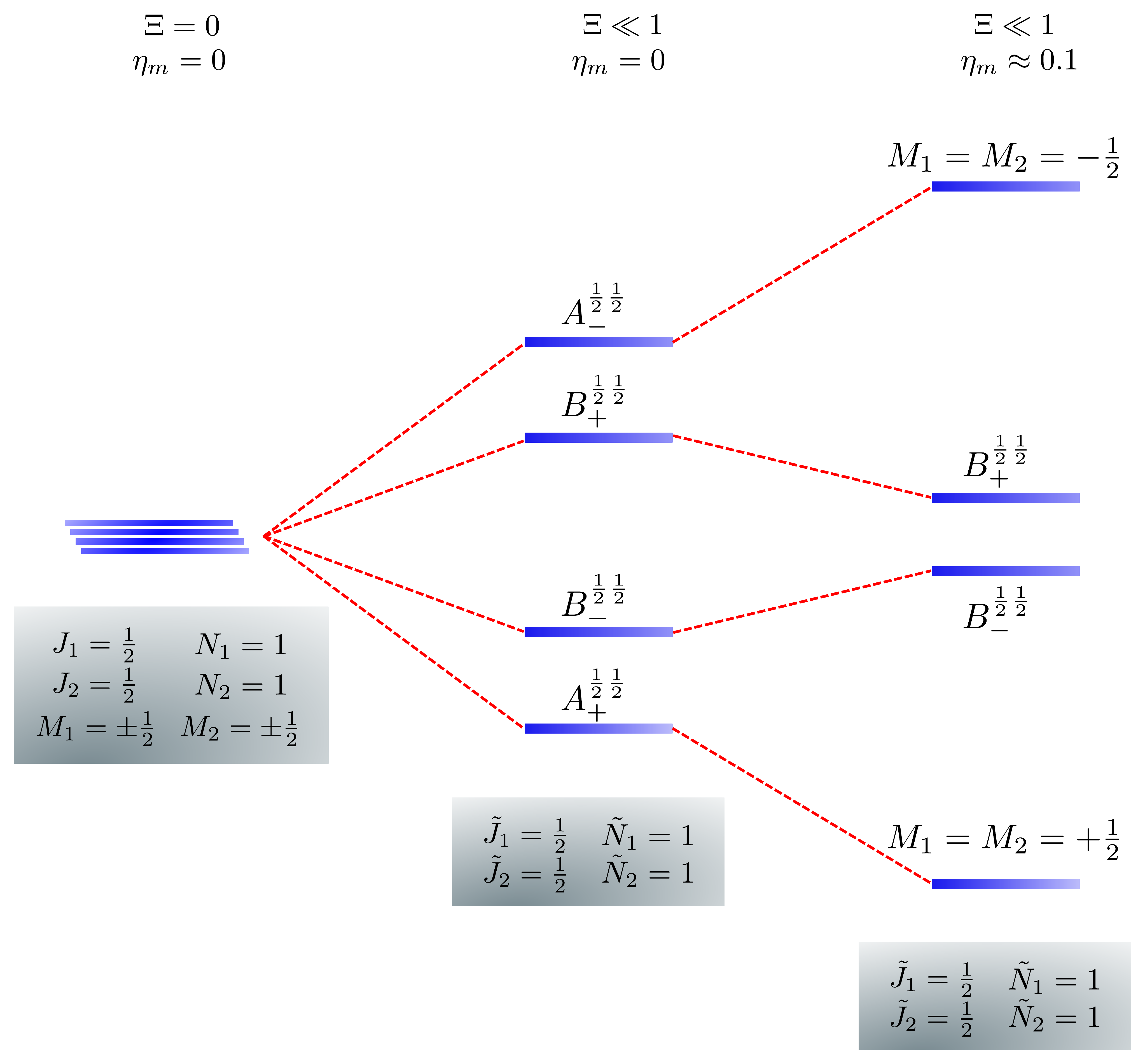}
  \caption{Correlation diagram involving the $\tilde{J}_1=\frac{1}{2}, \tilde{N}_1=1, \tilde{J}_2=\frac{1}{2}, \tilde{N}_2=1$ pair-eigenstates. The eigenstates, labelled in accordance with Table \ref{tbl:statelable},  are degenerate in the absence of the electric dipole-dipole interaction but their degeneracy is lifted when $\Xi\neq 0$. $A^{\frac{1}{2}\frac{1}{2}}_+$ and $A^{\frac{1}{2}\frac{1}{2}}_-$ states adiabatically transform into $M_1=M_2=\frac{1}{2}$ and $M_1=M_2=-\frac{1}{2}$ eigenstates, respectively, when the magnetic field is applied. Note that the $B$ states maintain their entanglement throughout.}
    \label{fig:cordiag3}
\end{figure}
 
The  pair-eigenstates formed as a result of the electric dipole-dipole interaction can no longer be factored into products of individual molecular eigenstates, as was the case above in Eq. (\ref{eqn:statedcmp}), and, moreover, even $M_1$ and $M_2$ cease to be good quantum numbers. Figures \ref{fig:cordiag1}-\ref{fig:cordiag3} show  correlation diagrams between the individual molecular eigenstates in the absence of the magnetic field ($\Xi=0$, $\eta_m=0$) and the pair-eigenstates created by the electric dipole-dipole interaction ($\Xi \neq 0$) without ($\eta_m=0$) and with ($\eta_m \neq 0$) the magnetic field for the three lowest sets of pair-eigenstates. 

In the absence of the magnetic field and the electric dipole-dipole interaction, the pair-eigenstates are degenerate in $M_1$ and $M_2$ for any given set of $J_1$, $N_1$, $J_2$ and $N_2$. Since $M_1=-J_1, -J_1+1, \dots J_1-1, J_1$ and $M_2=-J_2, -J_2+1, \dots J_2-1, J_2$, each such set is comprised of $(2J_1+1)(2J_2+1)$ degenerate states. The electric dipole-dipole interaction lifts the $M$-degeneracy as the pair-eigenstates are formed. In the absence of the magnetic field, the pair-eigenstates are equally-weighted linear combinations of the degenerate states of individual molecule with given $\pm M_1$ and $\pm M_2$. As indicated in the correlation diagrams of Figs. \ref{fig:cordiag1}-\ref{fig:cordiag3}, these linear combination states are formed irrespective of how small the value of $\Xi$ is. Every $\pm |M_1|$ and $\pm |M_2|$ set of degenerate states leads to the formation of four new pair-eigenstates.
  
  \begin{table}[h!]
  \centering
  \begin{tabular}{|c|c|}
  \hline
  Label & State \\
  \hline
  $A^{|M_1||M_2|}_+$ & $\Psi(+|M_1|, +|M_2|) + \Psi(-|M_1|, -|M_2|)$\\
  $A^{|M_1||M_2|}_-$ & $\Psi(+|M_1|, +|M_2|) - \Psi(-|M_1|, -|M_2|)$\\
  $B^{|M_1||M_2|}_+$ & $\Psi(+|M_1|, -|M_2|) + \Psi(-|M_1|, +|M_2|)$\\
  $B^{|M_1||M_2|}_-$ & $\Psi(+|M_1|, -|M_2|) - \Psi(-|M_1|, +|M_2|)$\\
  \hline
  \end{tabular}
  \caption{Pair-eigenstates -- and their labels -- comprised of two $^2\Sigma$ molecules in the presence of the electric dipole-dipole interaction. Note that these labels remain in place irrespective of whether the magnetic field is present.}
  \label{tbl:statelable}	
  \end{table}
  
Table \ref{tbl:statelable} shows the four possible states formed along with their respective labels. We label the states $A$ if the total angular momenta of the two molecules are parallel, i.e., the state is a linear combination of ($+|M_1|$, $+M_2$) and ($-|M_1|$, $-|M_2|$). The states are labelled $B$ if the total angular momenta of the two molecules are antiparallel. Note that the values of $|M_1|$ and $|M_2|$ are shown as superscripts whereas the subscripts $+$ and $-$ refer to whether the linear combination is symmetric or antisymmetric. 

As shown in Figs. \ref{fig:cordiag1}-\ref{fig:cordiag3}, in a magnetic field that lifts the $\pm M$ degeneracy, the $A$ states decouple into $+|M_1|, +|M_2|$ and $-|M_1|, -|M_2|$ states whereas the $B$ states do not (for as long as $M_1=M_2$). This is because in the $B$ states the two molecules have opposite projections of the angular momentum and the combinations  $\psi(+|M|, -|M|)$ and $\psi(-|M|,+|M|)$ are indistinguishable. This preserves the entanglement (the Bell-state character) of the pair-eigenstates even in the presence of a uniform magnetic field. However, the $B$ states decouple in a non-uniform magnetic field \cite{KarSharFri2016}. 

 \subsection{Pair-eigenstates in the presence of large dipole-dipole coupling, $\Xi \le 1$}
 \label{sec:strong}
 
  \begin{figure}[t!]
 \centering
  \includegraphics[width=\textwidth, height=0.7\textheight, keepaspectratio]{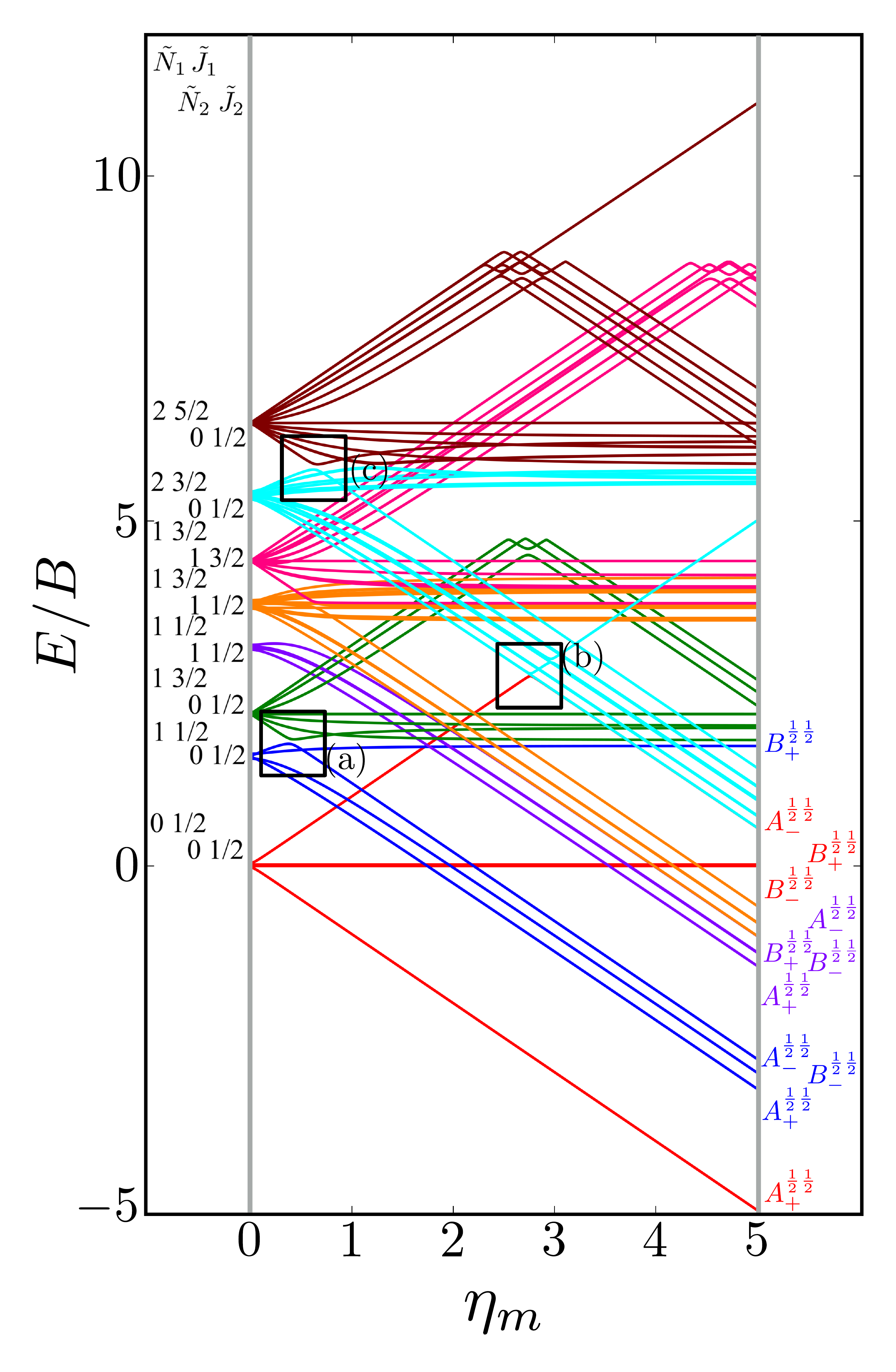}
  \caption{Dependence of the eigenenergies $E$ of the system of two polar paramagnetic $^2\Sigma$ molecules on the magnetic field strength parameter $\eta_m$ in the presence of the electric dipole-dipole interaction ($\Xi=10^{-1}$). The eigenenergies are measured in terms of the rotational constant $B$. The avoided crossings formed due to electric dipole-dipole interaction are highlighted by the black boxes. Cf. Fig. \ref{fig:novdd}.}
  \label{fig:highvdd}
\end{figure}  

 \begin{figure}[htp!]
 \centering
  \includegraphics[width=\textwidth, height=0.6\textheight, keepaspectratio]{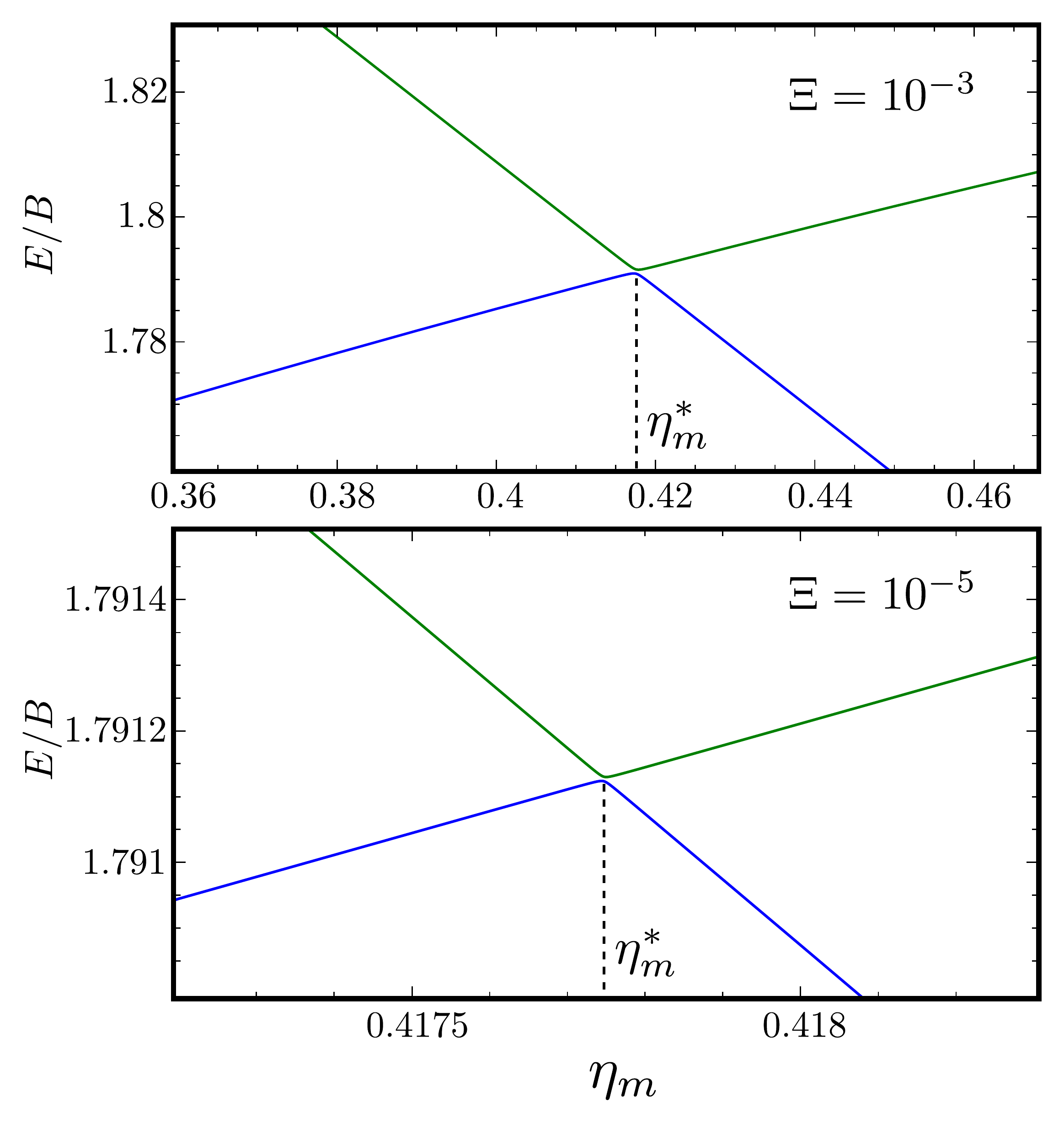}
  \caption{Zoomed-in plot of the first avoided crossing highlighted  by box (a) in Fig. \ref{fig:highvdd} for $\Xi=10^{-3}$ (upper panel) and for $\Xi=10^{-5}$ (lower panel). The position of the avoided crossing is marked by the value of the magnetic interaction parameter $\eta_m^*$.}
  \label{fig:zoom1}
\end{figure}  
 
In order to make the effect of the electric dipole-dipole interaction on the structure of the pair-eigenenergy levels more apparent,  we  increased the value of the coupling interaction parameter $\Xi$ to an unrealistically high value of 0.1, see Figure \ref{fig:highvdd}. Each set of pair-eigenstates with the same $\tilde{J}_1, \tilde{N}_1, \tilde{J}_2$ and $\tilde{N}_2$ are shown in the same colour. Since $M_1$ and $M_2$ are mixed, see Subsection \ref{sec:weak}, the eigenstates are labeled according to the system defined in Table \ref{tbl:statelable}. 

We see that avoided crossings  (highlighted by the black boxes) are formed for  pair-eigenstates comprised of individual states that meet the selection rules  $\Delta J_{i}=0,\pm 1$, $\Delta N_{i}=0,\pm 2$, and $\Delta M_{i}=0,\pm 1$.  These selection rules follow from the properties of the electric dipole-dipole operator, cf. Eq. (\ref{eqn:dip-dip}).  

Figure \ref{fig:zoom1} shows the first avoided crossing, highlighted by box (a) in Fig. (\ref{fig:highvdd}), for $\Xi=10^{-3}$ (upper panel) and $\Xi=10^{-5}$ (lower panel), illustrating the effect of increasing the value of $\Xi$.  The smaller the value of $\Xi$, the greater the zoom required in order to visualise the avoided crossing. 
 
 \subsection{Mutual alignment of the coupled rotors}
 \label{sec:alignment}
 
The alignment and orientation of the two-molecule system is characterised, respectively, by the expectation values of the pairwise alignment cosine $\cos\theta_1\cos\theta_2$ and pairwise orientation cosine $\cos^2\theta_1\cos^2\theta_2$ operators, see also  \cite{LemeshkoPrA2011,FriedrichMP2012}. The requisite matrix elements for calculating the pairwise cosines are listed in Appendix \ref{app:pwelements}. 
 
Figure \ref{fig:zoomcosine} shows the expectation values of the pairwise orientation and alignment cosines of the $\tilde{J}_1=\frac{1}{2}, \tilde{N}_1=1, B_+^{\frac{1}{2}\frac{1}{2}}$ state (blue curve in Fig. \ref{fig:highvdd}) with the $\tilde{J}_1=\frac{3}{2}, \tilde{N}_1=1, A_-^{\frac{3}{2}\frac{1}{2}}$ state (green curve in Fig. \ref{fig:highvdd}) at the avoided crossing for $\Xi=10^{-5}$ (lower panel) and $\Xi=10^{-3}$ (upper panel) at $\eta_m \approx 0.41775$. Note that these states are not oriented but there is a sudden change in alignment of the two molecules at the avoided crossing. 

As noted in our earlier work \cite{PCCP2000FriHer,JCP2000BoFri}, a small electric field can orient polar paramagnetic molecules in the presence of a magnetic field by virtue of the electric dipole coupling of the Zeeman levels. A similar effect is expected to arise for two polar paramagnetic molecules in a magnetic field due to coupling of their Zeeman levels by the electric dipole-dipole interaction, resulting in their mutual orientation. However, as shown in Fig.  \ref{fig:zoomcosine} (dashed line), the mutual orientation comes to naught. As detailed in Section \ref{sec:weak}, this is because the pair-eigenstates  are equally weighted linear combination of states with opposing  angular momentum projections on the space fixed $Z$ axis. In other words, the linear combinations entail indistinguishable pair-eigenstates of types $|\uparrow\downarrow\rangle$ and $|\downarrow\uparrow\rangle$. 

However, the molecules are mutually aligned by the electric dipole-dipole coupling, see Figure \ref{fig:zoomcosine}. 
 
 \begin{figure}[htp!]
 \centering
  \includegraphics[width=\textwidth, height=0.7\textheight, keepaspectratio]{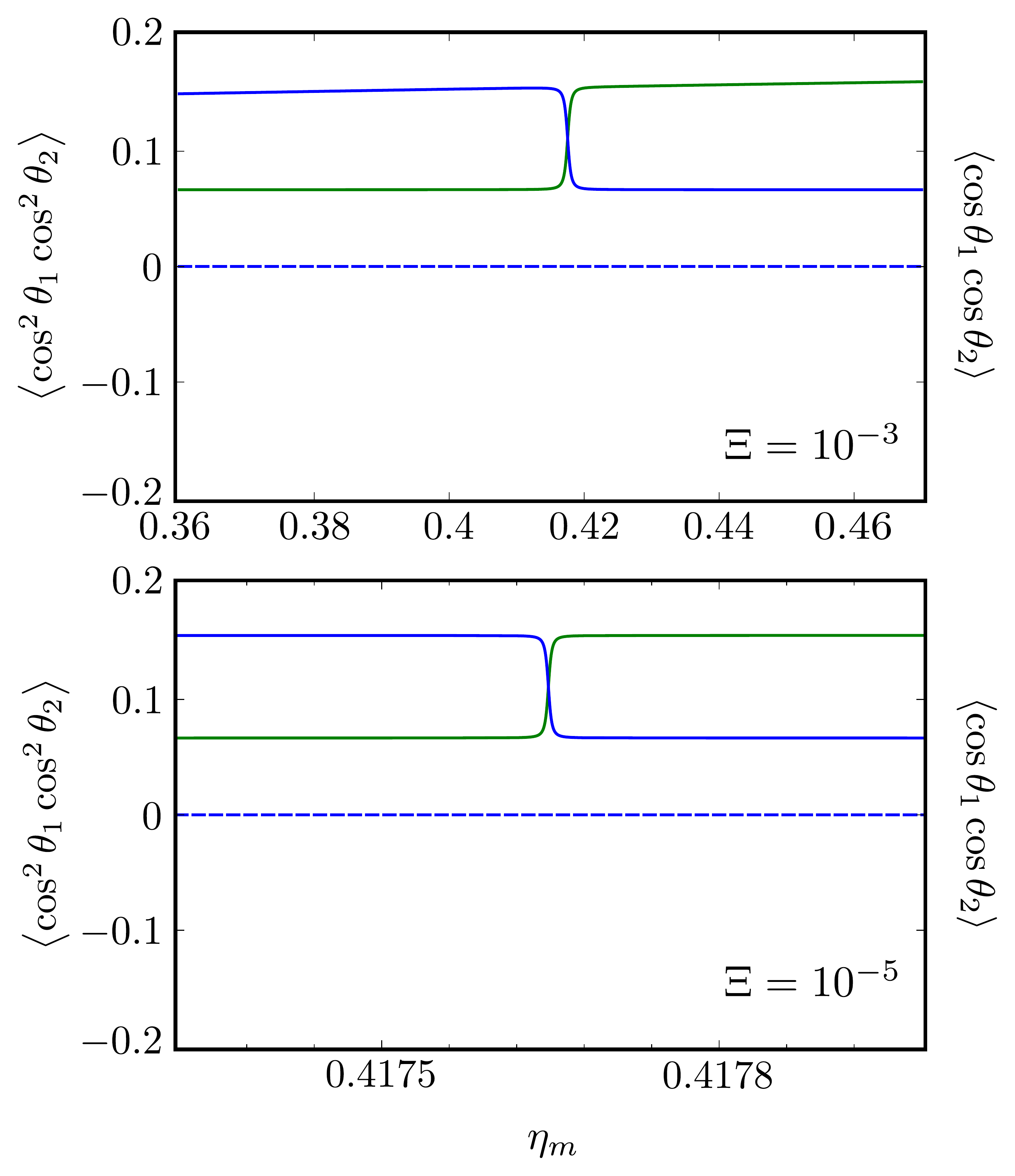}
  \caption{Pairwise alignment and orientation cosines of  two polar  $^2\Sigma$ molecules near the avoided crossing shown by a box (a) in Fig. \ref{fig:highvdd} as a function of the magnetic field strength parameter $\eta_m$ for electric dipole-dipole interaction $\Xi=10^{-3}$ (upper panel) and $\Xi=10^{-5}$ (lower panel).}
  \label{fig:zoomcosine}
 \end{figure}

 \begin{figure}[htp!]
 \centering
  \includegraphics[width=\textwidth, height=0.6\textheight, keepaspectratio]{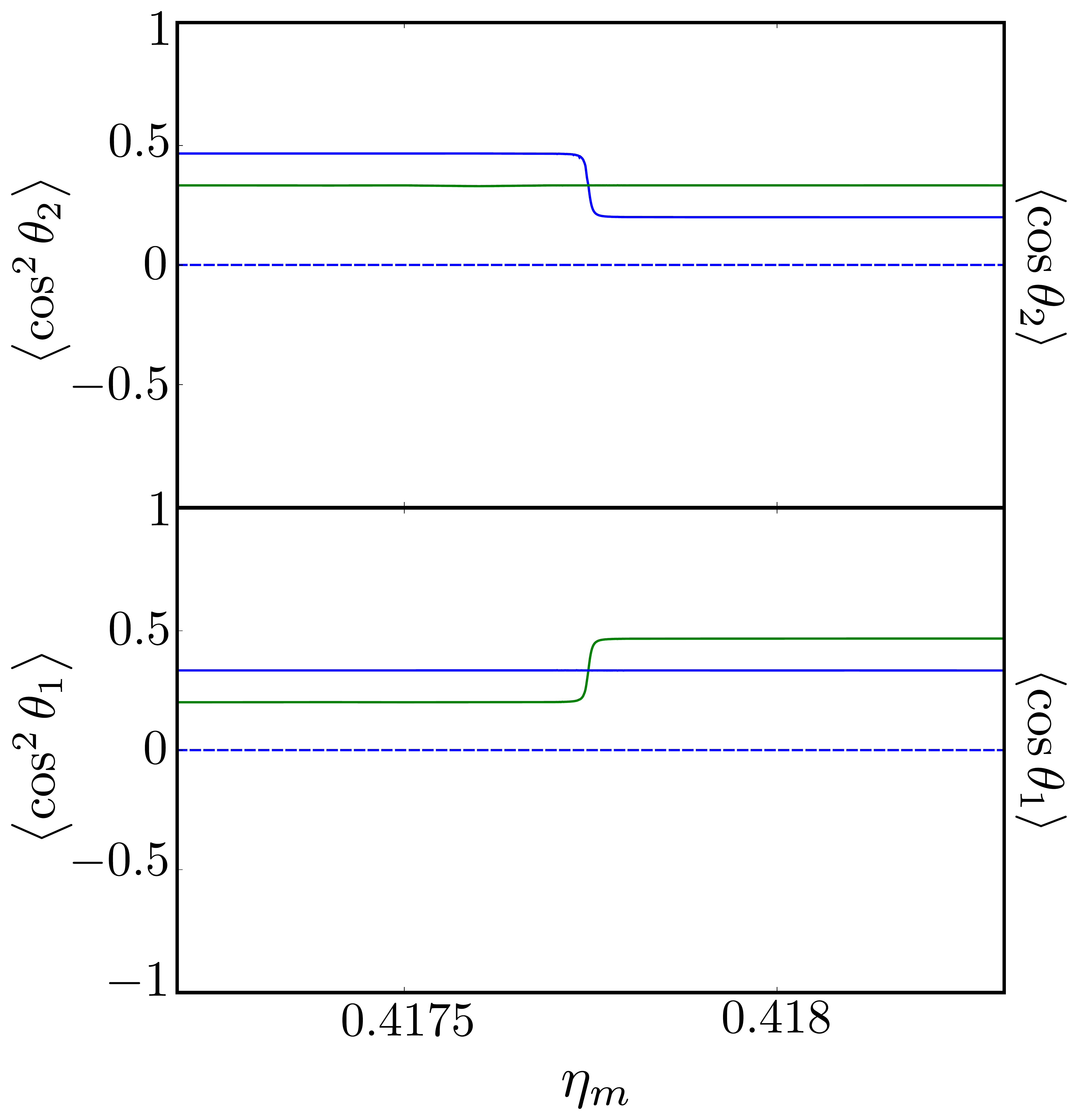}
  \caption{Individual alignment and orientation cosines of polar $^2\Sigma$ molecules 1 and 2 near the avoided crossing shown by a box (a) in Fig. \ref{fig:highvdd} as a function of the magnetic field strength parameter $\eta_m$ for electric dipole-dipole interaction $\Xi=10^{-5}$.}
  \label{fig:combined_alignment}
 \end{figure}
 
Figure \ref{fig:combined_alignment} shows the corresponding individual orientation and alignment cosines of  $\tilde{J}_1=\frac{1}{2}, \tilde{N}_1=1, B_+^{\frac{1}{2}\frac{1}{2}}$ state (blue curve in Fig. \ref{fig:highvdd}) with the $\tilde{J}_1=\frac{3}{2}, \tilde{N}_1=1, A_-^{\frac{3}{2}\frac{1}{2}}$ state (green curve in Fig. \ref{fig:highvdd}). The colour coding is the same as in Figs. \ref{fig:highvdd} and \ref{fig:zoomcosine}. We see that the coupling near the avoided crossing changes the alignment of one of the two molecules, which leads to a change in the mutual alignment shown in Fig. \ref{fig:zoomcosine}. For state $\tilde{J}_1=\frac{1}{2}, \tilde{N}_1=1, B_+^{\frac{1}{2}\frac{1}{2}}$ (blue curve), the alignment of molecule 1 remains constant but the alignment of molecule 2 decreases at the avoided crossing while for the state $\tilde{J}_1=\frac{3}{2}, \tilde{N}_1=1, A_-^{\frac{3}{2}\frac{1}{2}}$ the alignment of molecule 1 increases but the alignment of molecule 2 remains constant at the avoided crossing. The relationship between the individual alignment cosines of molecules 1 and 2 and the mutual alignment cosine (for given crossing states) illustrated in Figs. \ref{fig:zoomcosine} and \ref{fig:combined_alignment} is rendered by the two-state model below. 

 \subsection{Analytic model of pairwise alignment}
 \label{sec:analm}
 
For $\Xi \ll \eta_m$, the eigenproperties of two interacting eigenstates which cross in the purely magnetic case but form an avoided crossing in the presence of the electric dipole-dipole interaction can be calculated using a two-state model. This model makes use of the two Zeeman eigenfunctions in the absence of electric dipole-dipole coupling as the (unperturbed) basis functions. Thus
 \begin{equation}
 \begin{split}
  \left(H_1+H_2\right)\psi_a^{(0)}&=E_a^{(0)}\psi_a^{(0)} \\
  \left(H_1+H_2\right)\psi_b^{(0)}&=E_b^{(0)}\psi_b^{(0)} 
 \end{split}
 \end{equation}
where $H_1$ and $H_2$ are given by Eq. (\ref{eqn:hami}) and  $\psi_a^{(0)}\equiv\psi_a(\Xi=0)$ $\&$ $\psi_b^{(0)}\equiv\psi_b(\Xi=0)$ and the subscripts $a$ and $b$ pertain to the two states considered. In the absence of electric dipole-dipole interaction, these eigenfunctions are just a direct product of the eigenfunctions of individual molecules $i=1,2$,
 \begin{equation}
  \begin{split}
   \psi_a^{(0)}&=\phi_{1a}^{(0)}\phi_{2a}^{(0)}\\
   \psi_b^{(0)}&=\phi_{1b}^{(0)}\phi_{2b}^{(0)}
  \end{split}
 \end{equation}
where $\phi_i$ are the eigenfucntions of Hamiltonian (\ref{eqn:hami}) of molecules $i=1,2$. The eigenenergies of the pair-eigenstates $a$ and $b$ in the presence of the electric dipole-dipole coupling are then given by
 \begin{equation}
  \begin{split}
   E_a=&E_a^{(0)}-2\Delta E^{(0)}\left(\Xi\right)^\frac{1}{3}\left(1-\sec\alpha\right) \\
   E_b=&E_b^{(0)}+2\Delta E^{(0)}\left(\Xi\right)^\frac{1}{3}\left(1-\sec\alpha\right)
  \end{split}
  \label{eqn:twostateenergy}
 \end{equation}
 and the corresponding eigenvectors are given by
  \begin{equation}
  \left[ 
  \begin{matrix}
   \psi_a \\
   \psi_b 
  \end{matrix} 
  \right] = \left[
  \begin{matrix}
   \cos\alpha  & \sin\alpha \\
   -\sin\alpha & \cos\alpha
  \end{matrix}
  \right]
  \left[
  \begin{matrix}
   \psi_a^{(0)} \\
   \psi_b^{(0)}
  \end{matrix}
  \right]
 \end{equation}
 with $\alpha$ the mixing angle 
 \begin{equation}
  \alpha=\frac{1}{2}\tan^{-1}\left(\frac{2H_{12}}{\Delta E^{(0)}}\right)
 \end{equation}
 where $\Delta E^{(0)}\equiv E^{(0)}_b-E^{(0)}_a$, $H_{12}$ is the electric dipole-dipole coupling matrix element between the two unperturbed states,
 \begin{equation}
  H_{12}=\left\langle\psi_a^{(0)}\right|V_{dd}\left|\psi_b^{(0)}\right\rangle
 \end{equation}
 and $0^{\circ}\le \alpha \le 90^{\circ}$.
 Eq. (\ref{eqn:twostateenergy}) shows that the change in energy due to the electric dipole-dipole interaction is proportional to $\Xi^{\frac{1}{3}}$. Since $\Xi$ is inversely proportional to the cube of the distance between the molecules, cf. Eq. (\ref{eqn:Xi2}), we see that at large intermolecular separations the eigenenergies of the two molecule system due to electric dipole-dipole interaction vary as $r_{1,2}^{-1}$. 
 
Within the two-state model, the pairwise alignment cosine is given by
 \begin{equation}
 \label{eqn:aligncosine2}
 \begin{split}
  \left\langle \psi_{a,b}\left|\cos^2\theta_1\cos^2\theta_2\right|\psi_{a,b}\right\rangle= & \cos^2\alpha \left\langle\phi^{(0)}_{1a,b}\left|\cos^2\theta_1\right|\phi^{(0)}_{1a,b}\right\rangle\left\langle\phi^{(0)}_{2a,b}\left|\cos^2\theta_2\right|\phi^{(0)}_{2a,b}\right\rangle\\
  + & \sin^2\alpha \left\langle\phi^{(0)}_{1b,a}\left|\cos^2\theta_1\right|\phi^{(0)}_{1b,a}\right\rangle\left\langle\phi^{(0)}_{2b,a}\left|\cos^2\theta_2\right|\phi^{(0)}_{2b,a}\right\rangle\\
  \pm & \sin(2\alpha)\left\langle\phi^{(0)}_{1a,b}\left|\cos\theta_1\right|\phi^{(0)}_{1b,a}\right\rangle\left\langle\phi^{(0)}_{2a,b}\left|\cos\theta_2\right|\phi^{(0)}_{2b,a}\right\rangle
  \end{split}
 \end{equation}
 
Eq. (\ref{eqn:aligncosine2}) implies that for $\eta_m<\eta_m^*$ (with $\eta_m^*$ the magnetic field strength parameter corresponding to position of the avoided crossing), where $\alpha=0^{\circ}$, the pairwise alignment is a product of the alignment of states $a$ of molecules 1 and 2 and beyond the interaction region, where $\alpha=90^{\circ}$, the pairwise alignment is a product of the alignment of states $b$ of molecules 1 and 2. The pairwise alignment in the interaction (avoided crossing) region is a combination of the alignment of  states $a$ and $b$ plus an additional term which comes about due to the interaction. The interaction term reaches its maximum value at $\alpha=45^{\circ}$. 

We note that the maximum value of the pairwise alignment cosine is independent of the strength of the electric dipole-dipole coupling as long as $\Xi$ is nonzero. The pairwise alignment calculated from this model is quite accurate, within  $\pm 5\%$ of the exact result for $\Xi<10^{-3}$. Hence the model is quite useful, since typically  $\Xi\approx 10^{-5}$ for polar paramagnetic molecules at a distance of $500$ nm apart (for instance when trapped in an optical lattice). 

\section{Conclusion}
\label{sec:conc}
 
Our study of a composite system comprised of two polar $^2\Sigma$ molecules subject to a uniform magnetic field revealed that the electric dipole-dipole interaction that dominates the intermolecular potential between the two molecules mixes the molecules' $M$ states and in the process creates the maximally entangled Bell states. These are of two types, $A$ and $B$. While the entanglement of type $A$ states is destroyed by applying a magnetic field (which is tantamount to performing a Bell measurement on the system), the type $B$ states maintain their entanglement even in the presence of a uniform magnetic field. Only a non-uniform magnetic would destroy their entanglement as well. These features may find application in developing platforms for quantum computing with arrays of trapped molecules \cite{KarSharFri2016}.

Furthermore, we found that the intersecting Zeeman levels of the pair-eigenstates undergo avoided crossings if they obey  a set of selection rules imposed by the electric dipole-dipole operator: $\Delta J_{i}=0,\pm 1$, $\Delta N_{i}=0,\pm 2$, and $\Delta M_{i}=0,\pm 1$  with $J_{i}$, $N_{i}$ and $M_{i}$ the total, rotational and projection angular momentum quantum numbers of molecules $i=1,2$  in the absence of the electric dipole-dipole interaction. 

The two coupled rotors considered readily align each other in the absence of the magnetic field. Their mutual alignment depends on which rotational states of the two molecules are combined. A magnetic field modifies the mutual alignment in the vicinity of field strengths corresponding to the avoided crossings. An analytic model renders accurate values of the mutual alignment cosine for a wide range of dipole-dipole interaction and magnetic field strengths.
 
In our forthcoming work we plan to explore the effects of superimposed electric and non-resonant optical fields on the intermolecular  energy hypersurface, with special focus on the role of conical intersections of the Stark and Zeeman energy surfaces. We expect that this may suggest new ways of designing control fields for efficient and state-specific preparation of pair-states \cite{Sharmafriedrich2015}.

\begin{acknowledgments}

Discussions with Mallikarjun Karra as well as support by the Deutsche Forschungsgemeinschaft through grant FR 3319/3-1 are gratefully acknowledged.

\end{acknowledgments}

\newpage

\appendix

\section{Matrix elements of the electric dipole-dipole operator in the cross product basis set of the two molecules}
\label{app:dip_dip}

In the Hund's case (a) basis set of the two molecules, cf. Eq. (\ref{eqn:crossproduct})
\begin{equation}
 \ket{J_1 \Omega_1 M_1 S_1 \Sigma_1; J_2 \Omega_2 M_2 S_2 \Sigma_2}=\ket{J_1 \Omega_1 M_1 S_1 \Sigma_1}\otimes\ket{J_2 \Omega_2 M_2 S_2 \Sigma_2},
\end{equation}
 a general matrix element of $V_{d-d}$ becomes, cf. Eq. (\ref{eqn:Vdd}),
\begin{equation}
 \begin{split}
  \bra{J_1' \Omega_1' & M_1' S'_1 \Sigma'_1; J_2' \Omega_2' M_2' S'_2 \Sigma'_2} V_{d-d}\ket{J_1 \Omega_1 M_1 S_1 \Sigma_1; J_2 \Omega_2 M_2 S_2 \Sigma_2} \\
  = & -\sqrt{30}~\overline{\Xi} \sum_{\nu\,\lambda}\wjm{1}{1}{2}{\nu}{\lambda}{-\nu-\lambda}\wigmtx{2}{\nu+\lambda}{0}{\phi,\theta,\chi}A_1(\nu)A_2(\lambda) \delta_{S'_1 S_1} \delta_{S'_2 S_2} \delta_{\Sigma'_1 \Sigma_1} \delta_{\Sigma'_2 \Sigma_2}
 \end{split}
\end{equation}

where
\begin{equation}
 \label{eqn:A1}
 A_1(\nu)=\bra{J_1' \Omega_1' M_1'}\wigmtx{1}{-\nu}{0}{\phi_1,\theta_1,\chi_1}\ket{J_1 \Omega_1 M_1}
\end{equation}

\begin{equation}
 A_2(\lambda)=\bra{J_2' \Omega_2' M_2'}\wigmtx{1}{-\lambda}{0}{\phi_2,\theta_2,\chi_2}\ket{J_2 \Omega_2 M_2}
\end{equation}

Above and below we make use of the Wigner 3-J symbols instead of the  Clebsh-Gordon coefficients,
\begin{equation}
 \label{eqn:Cto3jm}
 C(j_1, j_2, j_3; m_1, m_2, m_3)=\left(-1\right)^{j_1-j_2+m_3}\sqrt{2j_3+1}\wjm{j_1}{j_2}{j_3}{m_1}{m_2}{-m_3}
\end{equation}
as well as of the identities
\begin{equation}
 \wigmtx{J}{M}{\Omega}{\omega}^\dagger=\left(-1\right)^{M-\Omega}\wigmtx{J}{-M}{-\Omega}{\omega},
\end{equation}

\begin{equation}
 \label{eqn:brasym}
 \bra{J \Omega M}=\left(\frac{2J+1}{8\pi^2}\right)^{\frac{1}{2}}\wigmtx{J}{M}{\Omega}{\omega},
\end{equation}
and
\begin{equation}
 \label{eqn:ketsym}
 \ket{J \Omega M}=(-1)^{M-\Omega}\left(\frac{2J+1}{8\pi^2}\right)^{\frac{1}{2}}\wigmtx{J}{-M}{-\Omega}{\omega}
\end{equation}
 where we abbreviated  $\left(\phi, \theta, \chi\right)$ as $\left(\omega\right)$.

From eqs. \eqref{eqn:A1}, \eqref{eqn:brasym}, and \eqref{eqn:ketsym}  we then obtain:
\begin{equation}
\label{A1}
 \begin{split}
 A_1(\nu)=&\bra{J_1' \Omega_1' M_1'}\wigmtx{1}{-\nu}{0}{\omega_1}\ket{J_1 \Omega_1 M_1} \\
   =&\left(\frac{2J_1'+1}{8\pi^2}\right)^{\frac{1}{2}}\left(\frac{2J_1+1}{8\pi^2}\right)^{\frac{1}{2}}\int\mathrm{d}\omega_1 \, \wigmtx{J_1'}{M_1'}{\Omega_1'}{\omega_1}\wigmtx{1}{-\nu}{0}{\omega_1}\wigmtx{J_1}{M_1}{\Omega_1}{\omega_1} 
 \end{split}
\end{equation}
and 
\begin{equation}
\label{A2}
 \begin{split}
 A_2(\nu)=&\bra{J_2' \Omega_2' M_2'}\wigmtx{1}{-\nu}{0}{\omega_1}\ket{J_2 \Omega_2 M_2} \\
   =&\left(\frac{2J_2'+1}{8\pi^2}\right)^{\frac{1}{2}}\left(\frac{2J_2+1}{8\pi^2}\right)^{\frac{1}{2}}\int\mathrm{d}\omega_2 \, \wigmtx{J_2'}{M_2'}{\Omega_2'}{\omega_2}\wigmtx{1}{-\nu}{0}{\omega_2}\wigmtx{J_2}{M_2}{\Omega_2}{\omega_2} 
 \end{split}
\end{equation}

By making use of the ``triple product theorem,''
\begin{equation}
 \label{eqn:3Dintg}
 \begin{split}
  \int\mathrm{d}\omega \, \wigmtx{J_3}{M_3}{\Omega_3}{\omega} & \wigmtx{J_2}{M_2}{\Omega_2}{\omega}\wigmtx{J_1}{M_1}{\Omega_1}{\omega} \\
   & = 8\pi^2 \wjm{J_1}{J_2}{J_3}{M_1}{M_2}{M_3}\wjm{J_1}{J_2}{J_3}{\Omega_1}{\Omega_2}{\Omega_3},
 \end{split}
\end{equation}
eqs. (\ref{A1}) and (\ref{A2}) reduce to
\begin{equation}
 A_1(\nu)=\left(2J_1'+1\right)^\frac{1}{2}\left(2J_1+1\right)^\frac{1}{2}\wjm{J_1'}{1}{J_1}{M_1'}{-\nu}{-M_1}\wjm{J_1'}{1}{J_1}{\Omega_1'}{0}{-\Omega_1}
\end{equation}
and
\begin{equation}
 A_2(\lambda)=\left(2J_2'+1\right)^\frac{1}{2}\left(2J_2+1\right)^\frac{1}{2}\wjm{J_2'}{1}{J_2}{M_2'}{-\lambda}{-M_2}\wjm{J_2'}{1}{J_2}{\Omega_2'}{0}{-\Omega_2}
\end{equation}
and so the complete electric dipole-dipole matrix element becomes:
\begin{equation}
\label{Vddmatrixelements}
 \begin{split}
  \bra{J_1^\prime \Omega_1^\prime M_1^\prime S_1^\prime \Sigma_1^\prime ; & J_2^\prime \Omega_2^\prime M_2^\prime S_2^\prime \Sigma_2^\prime}  V_{dd}\ket{J_1 \Omega_1 M_1 S_1 \Sigma_1 ; J_2 \Omega_2 M_2 S_2 \Sigma_2}  \\
  = & -\sqrt{30} \, \overline{\Xi} \, \left[2J_1'+1\right]^\frac{1}{2}\left[2J_1+1\right]^\frac{1}{2}\left[2J_2'+1\right]^\frac{1}{2}\left[2J_2+1\right]^\frac{1}{2}\\
  & \times \wjm{J_1'}{1}{J_1}{\Omega_1'}{0}{\Omega_1}\wjm{J_2'}{1}{J_2}{\Omega_2'}{0}{\Omega_2}   \delta_{S_1^\prime S_1}\delta_{S_2^\prime S_2} \delta_{\Sigma_1^\prime \Sigma_1} \delta_{\Sigma_2^\prime \Sigma_2}\\
    & \times \, \sum_{\nu\,\lambda} \wjm{1}{1}{2}{\nu}{\lambda}{-\nu-\lambda}\wigmtx{2}{\nu+\lambda}{0}{\phi,\theta,\chi}\wjm{J_1'}{1}{J_1}{M_1'}{-\nu}{M_1}\wjm{J_2'}{1}{J_2}{M_2'}{-\lambda}{M_2}    
 \end{split}
\end{equation}

The various mathematical identities used in this derivation are taken from Ref. \cite{Varshalovich2008}.

\section{Matrix elements of the pairwise alignment  cosine in the cross product basis set of the two molecules}
\label{app:pwelements}

The matrix element of the pairwise orientation cosine in  the cross product Hund's case (a) basis set of the two molecules is given by
 \begin{equation}
  \begin{split}
  \bra{J_1^\prime \Omega_1^\prime M_1^\prime S_1^\prime \Sigma_1^\prime & J_2^\prime \Omega_2^\prime M_2^\prime S_2^\prime \Sigma_2^\prime}  \cos\theta_1\cos\theta_2\ket{J_1 \Omega_1 M_1 S_1 \Sigma_1 J_2 \Omega_2 M_2 S_2 \Sigma_2}  \\
  & \bra{J_1^\prime \Omega_1^\prime M_1^\prime} \cos\theta_1 \ket{J_1 \Omega_1 M_1}\bra{J_2^\prime \Omega_2^\prime M_2^\prime} \cos\theta_2 \ket{J_2 \Omega_2 M_2} \delta_{S_1^\prime S_1}\delta_{S_2^\prime S_2} \delta_{\Sigma_1^\prime \Sigma_1} \delta_{\Sigma_2^\prime \Sigma_2}\\
  \end{split}
 \end{equation}
 and the matrix element of the  pairwise alignment cosine in the cross product Hund's case (a) basis set of the two molecules is
  \begin{equation}
  \begin{split}
  \bra{J_1^\prime \Omega_1^\prime M_1^\prime S_1^\prime \Sigma_1^\prime & J_2^\prime \Omega_2^\prime M_2^\prime S_2^\prime \Sigma_2^\prime}  \cos^2\theta_1\cos^2\theta_2\ket{J_1 \Omega_1 M_1 S_1 \Sigma_1 J_2 \Omega_2 M_2 S_2 \Sigma_2}  \\
  & \bra{J_1^\prime \Omega_1^\prime M_1^\prime} \cos^2\theta_1 \ket{J_1 \Omega_1 M_1}\bra{J_2^\prime \Omega_2^\prime M_2^\prime} \cos^2\theta_2 \ket{J_2 \Omega_2 M_2} \delta_{S_1^\prime S_1}\delta_{S_2^\prime S_2} \delta_{\Sigma_1^\prime \Sigma_1} \delta_{\Sigma_2^\prime \Sigma_2}\\
  \end{split}
 \end{equation}
 The matrix elements of ${\bf S}_Z$, $\cos\theta$ and $\cos^2\theta$ in the symmetric top basis set are listed in Ref. \cite{Sharmafriedrich2015}.
\bibliographystyle{rsc}
 \bibliography{Dipoledipolebib}
\end{document}